\documentclass[aps,prl,reprint,groupedaddress]{revtex4-1} %reprint
\usepackage[dvipsnames]{xcolor}
\usepackage[colorlinks=true,urlcolor=blue,citecolor=purple,linkcolor=red]{hyperref}
\usepackage{amssymb,amsmath,amsfonts}
\usepackage{epsfig}
\usepackage{graphicx}
\usepackage{epstopdf}
\def\clock{{\count0=\time
           \divide\count0 60
           \ifnum\count0<10 0\fi\the\count0
           \multiply\count0 -60 \advance\count0 \time
           :\ifnum\count0<10 0\fi \the\count0
         }}
\newcommand{\timestamp}{{\small\vbox{\hbox{\tt\jobname.tex}
\hbox{\the\day/\the\month/\the\year, \clock}}}}

\newcommand{\CN}{\mathcal{N}}

\newcommand{\C}{\mathbb{C}}
\newcommand{\R}{\mathbb{R}}

\newcommand{\nn}{\nonumber}

\newcommand{\spa}{\ , \ \ }

\newcommand{\be}{\begin{eqnarray}}
\newcommand{\ee}{\end{eqnarray}}
\newcommand{\beq}{\begin{eqnarray}}
\newcommand{\eeq}{\end{eqnarray}}

\newcommand{\beqa}{\begin{eqnarray}}
\newcommand{\eeqa}{\end{eqnarray}}

\newcommand{\ds}{\displaystyle}

\let\oldsqrt\sqrt
\def\sqrt{\mathpalette\DHLhksqrt}
\def\DHLhksqrt#1#2{%
\setbox0=\hbox{$#1\oldsqrt{#2\,}$}\dimen0=\ht0
\advance\dimen0-0.2\ht0
\setbox2=\hbox{\vrule height\ht0 depth -\dimen0}%
{\box0\lower0.4pt\box2}}

\newcommand{\ads}{\mbox{AdS}}

\begin{document}

\title{Non-Relativistic Strings and Limits of the AdS/CFT Correspondence}

\author{Troels Harmark$^1$, Jelle Hartong$^2$, Niels A. Obers$^1$}
\email[]{harmark@nbi.ku.dk, j.hartong@uva.nl, obers@nbi.ku.dk}
%\thanks{}
\affiliation{$^1$The Niels Bohr Institute, University of Copenhagen}
\address{Blegdamsvej 17, 2100 Copenhagen \O, Denmark}
\affiliation{$^2$Institute for Theoretical Physics and Delta Institute for Theoretical Physics, University of Amsterdam\\
Science Park 904, 1098 XH Amsterdam, The Netherlands}


%\date{\today}

\begin{abstract}

Using target space null reduction of the Polyakov action we find a novel covariant action for strings moving in a torsional Newton--Cartan geometry. 
Sending the string tension to zero while rescaling the Newton--Cartan clock 1-form, so as to keep the string action finite, we obtain a non-relativistic string moving in a new type of non-Lorentzian geometry that we call $U(1)$-Galilean geometry. We apply this to strings on $AdS_5 \times S^5$ for which we show that the zero tension limit is realized by the Spin Matrix theory limits of the AdS/CFT correspondence. This is closely related to limits of spin chains studied in connection to integrability in AdS/CFT. The simplest example gives a covariant version of the Landau-Lifshitz sigma-model.

\end{abstract}

%\keywords{}

\maketitle

\noindent\textbf{Introduction} 

Non-Lorentzian geometry has appeared in recent years in a wide variety of settings such as non-AdS holography \cite{Christensen:2013lma,Hartong:2014oma}, effective actions of non-relativistic field theories including those relevant for the fractional quantum Hall effect \cite{Son:2013rqa,Geracie:2014nka,Jensen:2014aia,Hartong:2014pma} and gravity theories
with non-relativistic local symmetries such as Ho\v rava--Lifshitz gravity and Chern--Simons gravity theories on non-relativistic algebras 
\cite{Hartong:2015zia,Bergshoeff:2016lwr,Hartong:2016yrf}.  

By non-Lorentzian geometry we mean a manifold that is locally flat in the sense of a kinematical principle of relativity that is different from Einstein's equivalence principle. Examples are Newton--Cartan and Carrollian geometries whose tangent space structure is dictated by the Bargmann (centrally extended Galilei) and Carroll (zero speed of light contraction of Poincar\'e) algebras.

There is considerable literature on non-relativistic strings, see e.g. \cite{Gomis:2000bd,Kruczenski:2003gt,Andringa:2012uz,Batlle:2016iel}. Of particular relevance for us will be the non-relativistic string spectra and associated sigma-models, such as the Landau--Lifshitz model, observed before in the AdS/CFT context \cite{Kruczenski:2003gt}. Besides the stringy Newton-Cartan geometry found in \cite{Andringa:2012uz}, these works are non-covariant with regards to the world-sheet and target space geometry. A natural question is thus to what extent non-Lorentzian geometries are important for sigma-models of non-relativistic strings.  

In this letter we show that target space null-reduction of the Polyakov action leads to a novel covariant action for the propagation of non-relativistic strings in a (torsional) Newton-Cartan target space. Furthermore, we uncover that taking a second non-relativistic limit, that affects both the target space and the world-sheet, leads to a new class of sigma-models that describes non-relativistic strings moving in a novel non-Lorentzian geometry that we refer to as $U(1)$-Galilean geometry.

Remarkably, we show that for a string on $AdS_5 \times S^5$, the second non-relativistic limit corresponds to the Spin Matrix theory limits of the AdS/CFT correspondence. Spin Matrix theories are quantum mechanical theories that arise as limits of $\CN=4$ SYM on $\R\times S^3$ \cite{Harmark:2014mpa}. Given a unitarity bound $E\geq J$ of $\CN=4$ SYM, where $J$ is a linear combination of commuting angular momenta and $R$-charges such that states with $E=J$ are supersymmetric, one sends $E-J$ and the 't Hooft coupling $\lambda = 4\pi g_s N$ to zero, keeping the ratio 
$(E-J)/\lambda$ and $N$ fixed. It is clear  from the relativistic magnon dispersion relation \cite{Beisert:2005tm} $E-J = \sqrt{1 + \frac{\lambda}{\pi}  \sin^2 \frac{p}{2}}-1$ that a non-relativistic spectrum is obtained in this limit. We show that the SMT limits \cite{Harmark:2008gm, Harmark:2014mpa} on the string theory side correspond to our double (target space/world-sheet) non-relativistic limit. 

The SMT limits are closely related to limits of strings on $AdS_5 \times S^5$ considered in connection with integrability of the AdS/CFT correspondence, starting with Kruczenski \cite{Kruczenski:2003gt}. The difference is that the Kruczenski limit does not decouple higher order terms in the string tension. However, the leading part of the sigma-model is the same as for SMT. 

From the SMT or Kruczenski limit one gets the well-known Landau-Lifshitz sigma-model in the simplest case. Other limits give similar sigma-models that also are classical limits of nearest-neighbor spin chains \cite{Kruczenski:2004kw,Hernandez:2004uw,Stefanski:2004cw,Bellucci:2004qr}. 
Using the results of this letter, these sigma-models can be made covariant, thus providing a new interpretation  in terms of non-relativistic string theory.

\noindent\textbf{Strings on torsional Newton--Cartan geometry}

The action of a non-relativistic particle moving in a torsional Newton--Cartan (TNC) geometry can be obtained by null reduction of the action of a relativistic massless particle \cite{Duval:1984cj,Duval:1990hj,Festuccia:2016caf}. Here we will do something similar for the target space null reduction of the Polyakov action for a relativistic string. 

Consider the Polyakov action, 
\begin{equation}
S=\int d^2\sigma\mathcal{L}=-\frac{T}{2}\int d^2\sigma\sqrt{-\gamma}\gamma^{\alpha\beta}g_{\alpha\beta}\,,
\end{equation}
with $g_{\alpha\beta}=\partial_\alpha X^{M}\partial_\beta X^{N}G_{MN}$ where $G_{MN}$ is the $d+2$ dimensional target space metric. Here $\partial_\alpha$ is the derivative with respect to the world-sheet coordinates $\sigma^\alpha$ with $\alpha=0,1$, and $T$ is the string tension. 
We consider closed strings hence $\sigma^1\sim\sigma^1+2\pi$. 
The Virasoro constraints are
\begin{equation}\label{eq:Virasoroconstraints}
g_{\alpha\beta}-\frac{1}{2}\gamma_{\alpha\beta}\gamma^{\gamma\delta}g_{\gamma\delta}=0\,.
\end{equation}

Assume that the target space has a null Killing vector $\partial_u$. The most general metric with this property is
\begin{equation}\label{eq:nullred}
G_{MN}dx^{M}dx^{N}=2\tau\left(du-m\right)+h_{\mu\nu}dx^\mu dx^\nu\,,
\end{equation}
where $\mu,\nu=0,1,...,d$, $M=(u,\mu)$ and $\tau=\tau_\mu dx^\mu$, $m=m_\mu dx^\mu$, $\text{det}\, h_{\mu\nu}=0$. The tensors $\tau_\mu$, $m_\mu$ and $h_{\mu\nu}$ are independent of $u$. This decomposition of the line element admits the following local symmetries
\begin{eqnarray}
\delta \tau_\mu & = & \mathcal{L}_\xi\tau_\mu\,,\qquad\delta m_\mu = \mathcal{L}_\xi m_\mu+\partial_\mu\sigma+\lambda_a e^a_\mu\,,\nonumber\\
\delta h_{\mu\nu} & = & \mathcal{L}_\xi h_{\mu\nu}+\tau_\mu\lambda_a e^a_\nu+\tau_\nu\lambda_a e^a_\mu\,,\label{eq:TNCsym}
\end{eqnarray}
where we defined $e_\mu^a$ via $h_{\mu\nu}=\delta_{ab}e^a_\mu e^b_\nu$ with $a=1,\ldots,d$. The transformation with parameter $\sigma$ is a $U(1)$ gauge transformation that acts on $u$ as $\delta u=\sigma$. The transformation with parameter $\lambda_a$ is known as a local Galilean or Milne boost. The Lie derivatives along $\xi^\mu$ correspond to the infinitesimal $d+1$ dimensional diffeomorphisms. The fields and transformations \eqref{eq:TNCsym} are those of torsional Newton--Cartan (TNC) geometry \cite{Andringa:2010it,Jensen:2014aia,Hartong:2014pma,Hartong:2015zia,Geracie:2015dea} in agreement with the known fact that null reductions give rise to TNC geometry \cite{Duval:1984cj,Duval:1990hj,Julia:1994bs,Christensen:2013rfa}.

So far we are still describing a relativistic string in a background with a null isometry. To turn this into a non-relativistic string moving in a TNC background we need to remove the field $X^u$ from the description. This is achieved by putting the momentum $P_u^\alpha$ along $u$, 
\begin{equation}\label{eq:momstringu}
P^\alpha_u=\frac{\partial\mathcal{L}}{\partial\left(\partial_\alpha X^u\right)}=-T\sqrt{-\gamma}\gamma^{\alpha\beta}\tau_\beta\,,
\end{equation}
on-shell, i.e.~imposing $\partial_\alpha P_u^\alpha=0$, here defining $\tau_\beta=\partial_\beta X^\mu\tau_\mu$ as the pullback of $\tau_\mu$. This requires considering $P_u^\alpha$ (as opposed to $\partial_\alpha X^u$) as an independent variable. We thus perform the Legendre transformation
\begin{equation}
\hat{\mathcal{L}}=\mathcal{L}-P^\alpha_u\partial_\alpha X^u\,,
\end{equation}
where $\hat{\mathcal{L}}$ is the Lagrangian for the remaining embedding coordinates $X^\mu$ whose dependence on $P_u^\alpha$ is such that 
\begin{equation}\label{eq:Peom}
\frac{\partial\hat{\mathcal{L}}}{\partial P_u^\alpha}=-\partial_\alpha X^u\,.
\end{equation}

We will use \eqref{eq:momstringu} to solve for $\gamma_{\alpha\beta}$ in terms of $P_u^\alpha$ and $\tau_\alpha$. The solution to \eqref{eq:momstringu} can be written as
\begin{equation}
\sqrt{-\gamma}\gamma^{\alpha\beta}=e\left(-v^\alpha v^\beta+e^\alpha e^\beta\right)\,,
\end{equation}
where we defined $e=\text{det}(\tau_\alpha\,,e_\alpha)=\frac{1}{T}P_u^\alpha\tau_\alpha$ and
\begin{equation}
\label{esystem}
e_\alpha=\frac{e_{\alpha\beta}P^\beta_u}{T}\,, \quad v^\alpha=-\frac{P_u^\alpha}{P_u^\gamma\tau_\gamma}\,,\quad e^\alpha=-T\frac{e^{\alpha\beta}\tau_\beta}{P_u^\gamma\tau_\gamma}\,.
\end{equation}
Here $e_{\alpha\beta}$ and $e^{\alpha\beta}$ denote Levi-Civita symbols with $e^{01}=-e_{01}=1$. Together with $\tau_\alpha$ the vectors \eqref{esystem} form an orthonormal system: $v^\alpha\tau_\alpha=-1$, $v^\alpha e_\alpha=0$, $e^\alpha\tau_\alpha=0$ and $e^\alpha e_\alpha=1$. We assume that $P_u^\alpha\tau_\alpha\neq 0$.

The action associated with $\hat{\mathcal{L}}$ can be written as
\begin{equation}
\hat S =  \int d^2\sigma\hat{\mathcal{L}}= -\frac{T}{2}\int d^2\sigma e\left(-v^\alpha v^\beta+e^\alpha e^\beta\right)\bar h_{\alpha\beta}\\
\end{equation}
where $\bar h_{\alpha\beta}=\partial_\alpha X^\mu\partial_\beta X^\nu\bar h_{\mu\nu}$ with $\bar h_{\mu\nu}=h_{\mu\nu}-\tau_\mu m_\nu-\tau_\nu m_\mu$. Further $m_\alpha$ and $h_{\alpha\beta}$ are the pullbacks of $m_\mu$ and $h_{\mu\nu}$. From \eqref{eq:Peom} we obtain
\begin{equation}\label{eq:Virasoroform}
m_\alpha-\frac{1}{2}\tau_\alpha\left(e^\gamma e^\delta+v^\delta v^\gamma\right) h_{\gamma\delta} +e_\alpha  v^\gamma e^\delta h_{\gamma\delta} = \partial_\alpha X^u\,,
\end{equation}
which is equivalent to the Virasoro constraints \eqref{eq:Virasoroconstraints} for a string in the background with a null isometry \eqref{eq:nullred}. This follows from contracting 
\eqref{eq:Virasoroconstraints} with all combinations of $e^\alpha$ and $v^\alpha$. Furthermore from \eqref{eq:Peom} it follows that
\begin{equation}\label{eq:TNCconstraint}
\partial_\alpha\frac{\partial\hat{\mathcal{L}}}{\partial P_u^\beta}-\partial_\beta\frac{\partial\hat{\mathcal{L}}}{\partial P_u^\alpha}=0\,,
\end{equation}
which is independent of $X^u$.

We are now going to put $P_u^\alpha$ on-shell, i.e. impose $\partial_\alpha P_u^\alpha=0$ which is equivalent to setting $\partial_\alpha e_\beta-\partial_\beta e_\alpha=0$. We will write $P^\alpha_u=Te^{\alpha\beta}e_\beta$ where locally $e_\beta=\partial_\beta\eta$ and substitute this into the action $\hat S$. This leads to the following Lagrangian for $X^\mu$ and $\eta$,
\begin{equation}\label{eq:NRstringTNC}
\hat {\cal{L}}=T \left(-e^{\alpha\beta}m_\alpha\partial_\beta\eta + \frac{e^{\alpha\alpha'}e^{\beta\beta'}\left(\partial_{\alpha'}\eta\partial_{\beta'}\eta-\tau_{\alpha'}\tau_{\beta'}\right)}{2e^{\gamma\gamma'}\tau_{\gamma} \partial_{\gamma'}\eta}h_{\alpha\beta}\right) .
\end{equation}

The equation of motion of $\eta$ gives the constraint \eqref{eq:TNCconstraint}. The action \eqref{eq:NRstringTNC} is invariant under world-sheet diffeomorphisms $\delta X^\mu=\xi^\alpha\partial_\alpha X^\mu$ and $\delta\eta=\xi^\alpha\partial_\alpha\eta$ generated by $\xi^\alpha$, as well as under all local symmetries of the target space TNC geometry that are generated by $\sigma$ and $\lambda^a$ in \eqref{eq:TNCsym}. There can also be global symmetries generated by $K^\mu$ for those $\xi^\mu=K^\mu$ in \eqref{eq:TNCsym} for which $0=\delta \tau_\mu=\delta m_\mu=\delta h_{\mu\nu}$.

Assume that the target space clock 1-form $\tau$ is closed. Write this as $\tau_\mu=\partial_\mu X^0$. We can then choose the gauge $\sigma^0=\frac{2\pi T}{P}X^0$ and $\eta=\frac{P}{2\pi T}\sigma^1$ with $P=\int_0^{2\pi}P_u^0d\sigma^1$ the conserved total momentum. 
In this gauge the action \eqref{eq:NRstringTNC} on a flat TNC background with $m_\mu=0$, $\tau_\mu=\delta_\mu^0$ and $h_{\mu\nu}=\delta_{ab}\delta^a_\mu\delta^b_\nu$ reproduces the standard non-relativistic string action which has 1+1 dimensional world-sheet Poincar\'e symmetry \cite{Zwiebach:2004tj}. This latter action was also studied in \cite{Andringa:2012uz}. However the coupling to the target space geometry in \cite{Andringa:2012uz} involves a doubling of the fields $\tau_\mu$ and $m_\mu$ which we do not see here. It would be interesting to understand this difference.

\noindent\textbf{Non-relativistic sigma models from scaling limit} 

We will take a limit of $\hat S$ in which the tension $T$ goes to zero. In order to keep the action finite we compensate $T\rightarrow 0$ by rescaling the coupling to $\tau_\mu$. 
We can always write $\tau_\mu=N\partial_\mu F+\beta_\mu$ with $v^\mu\beta_\mu=v^\mu h_{\mu\nu}=0$  and $v^\mu\tau_\mu=-1$. If we rescale $F=c^2\tilde F$, $T=\tilde T/c$, $\eta=c\tilde\eta$ and send $c$ to infinity we obtain
\begin{eqnarray}
\hspace{-.5cm}\tilde S & = & -\tilde T \int d^2\sigma\left(e^{\alpha\beta}m_\alpha\partial_\beta\tilde\eta+\frac{e^{\alpha\alpha'}e^{\beta\beta'}\tilde\tau_{\alpha'}\tilde\tau_{\beta'}}{2e^{\gamma\gamma'}\tilde\tau_\gamma\partial_{\gamma'}\tilde\eta} h_{\alpha\beta}\right) ,\label{eq:tildeS}
\end{eqnarray}
where $\tilde\tau_\alpha=\partial_\alpha X^\mu\tilde\tau_\mu$ with $\tilde\tau_\mu=N\partial_\mu\tilde F$.

The resulting action $\tilde S$ has world-sheet diffeomorphism invariance $\delta X^\mu=\xi^\alpha\partial_\alpha X^\mu$ and $\delta\tilde\eta=\xi^\alpha\partial_\alpha\tilde\eta$. 
Assuming $\tilde{\tau}_\mu = \partial_\mu X^0 = \delta_\mu^0$, we can choose the gauge $\sigma^0 = \frac{4\pi^2 \tilde{T}^2}{P^2} X^0$ and $\tilde{\eta}=\frac{P}{2\pi \tilde{T}} \sigma^1$, obtaining 
%If we go to the gauge $\tilde\tau_0=\frac{P^2}{4\pi^2\tilde T^2}$, $\tilde\tau_1=0$ (assuming that $\tau_\mu=\partial_\mu X^0$) and $P_u^0=\frac{P}{2\pi}$, $P_u^1=0$ we obtain
\begin{eqnarray}
\label{eq:Stildegauge0}
\tilde S & = & -\frac{P}{2\pi} \int d^2\sigma\left(m_\mu \partial_0 X^\mu + \frac{1}{2}h_{\mu\nu} \partial_1 X^\mu \partial_1 X^\nu \right) . 
\end{eqnarray}
This is a non-relativistic world-sheet theory containing only first order time derivatives.
The equation of motion of $\tilde\eta$ gives the constraint
\begin{equation}
\partial_0m_1-\partial_1 m_0+\frac{1}{2}\partial_1 h_{11}  =0\,.
\end{equation}

The action \eqref{eq:tildeS} is invariant under local transformations that act on $\tilde\tau_\mu$, $m_\mu$ and $h_{\mu\nu}=\delta_{ab}e^a_\mu e^b_\nu$ as
\begin{equation}\label{eq:limitTNCsym}
\delta\tilde\tau_\mu=0\,,\quad\delta m_\mu = \partial_\mu\sigma\,,\quad\delta h_{\mu\nu} =2 \tilde\tau_{(\mu} e^a_{\nu)}\tilde\lambda_a\,.
\end{equation}
These transformations plus target space diffeomorphsims follow from \eqref{eq:TNCsym} if we set $\lambda_a=\tilde\lambda_a/c^2$, $\tau_\mu=c^2\tilde\tau_\mu+\beta_\mu$ and send $c$ to infinity. The action $\tilde S$ has a global symmetry generated by $K^\mu$ if the Lie derivatives along $K^\mu$ of $\tilde\tau_\mu$, $m_\mu$, $h_{\mu\nu}$ vanish up to the transformations \eqref{eq:limitTNCsym}. 

TNC geometry can be obtained by gauging the Bargmann algebra \cite{Andringa:2012uz,Hartong:2015zia}. The transformations \eqref{eq:TNCsym} follow from the Bargmann algebra $\{H, P_a, J_{ab}, G_a, N\}$ with $a=1,\ldots,d$ whose nonzero commutators are $\left[H,G_a\right]=P_a$ and $\left[P_a,G_b\right]=\delta_{ab}N$ where we left out the nonzero commutators with $J_{ab}$. The TNC fields can be assembled in the connection $\mathcal{A}_\mu=H\tau_\mu+P_a e^a_\mu+Nm_\mu+\ldots$, where we left out the connections associated with Galilean boosts $G_a$ and rotations $J_{ab}$. If we consider the transformation $\delta\mathcal{A}_\mu=\mathcal{L}_\xi \mathcal{A}_\mu+\partial_\mu\Sigma+\left[\mathcal{A}_\mu,\Sigma\right]$, where $\xi^\mu$ generates diffeomorphisms and where $\Sigma=N\sigma+G_a\lambda^a+\frac{1}{2}J_{ab}\lambda^{ab}$ we obtain all transformations of the TNC fields $\tau_\mu$, $m_\mu$ and $h_{\mu\nu}=\delta_{ab}e^a_\mu e^b_\nu$ in \eqref{eq:TNCsym}. If we rescale $H=c^2\tilde H$ and $G_a=c^{-2}\tilde G_a$ and send $c$ to infinity we find the Galilei algebra $\texttt{Gal}$ direct sum with a $U(1)$ generated by $N$, where $\texttt{Gal}$ is the Bargmann algebra with $N$ removed. In a similar way the local transformations of $\tilde\tau_\mu$, $m_\mu$ and $h_{\mu\nu}=\delta_{ab}e^a_\mu e^b_\nu$ can be obtained by gauging $\texttt{Gal}\oplus U(1)$ where $\tilde\tau_\mu$ is the connection associated with $\tilde H$, $e^a_\mu$ the connection associated with $P_a$ and $m_\mu$ the connection associated with $N$. The resulting geometry is what we call $U(1)$-Galilean geometry.

Interestingly, applying the same limit to the case of a massless relativistic particle leads to an action proportional to $\int d\lambda m_\mu\frac{dX^\mu}{d\lambda}$, so that a particle on a $U(1)$-Galilean geometry has no dynamics. We have thus found a geometry that is more naturally probed by strings than by particles.

\noindent\textbf{Limits of strings on AdS$_5\times S^5$}

We apply now the above scaling limit $c\rightarrow \infty$ to the case of strings on AdS$_5\times S^5$. As we shall see, the Spin Matrix theory (SMT) limits introduced in \cite{Harmark:2014mpa} are realizations of the scaling limit. 
Consider type IIB strings on AdS$_5\times S^5$ in the global patch with radius $R = (4\pi g_s N)^{1/4} l_s$ and five-form flux $N$ where $g_s$ is the string coupling and $l_s$ the string length. Introduce now the following six commuting charges, namely the energy $E$, the angular momenta $S_1$ and $S_2$ on the $S^3$ in AdS$_5$ and the angular momenta $J_1$, $J_2$ and $J_3$ on $S^5$. The unitarity bounds of $\CN=4$ are dual to BPS bounds 
$E \geq J$ where $J$ is a linear combination of the five angular momenta. Specifically, one has the five BPS bounds $E\geq J$ with $J= J_1+J_2$, $J=J_1+J_2+J_3$, $J=S_1+J_1+J_2$, $J=S_1+S_2+J_1$ or $J=S_1+S_2+J_1+J_2+J_3$. For a given BPS bound $E\geq J$ the SMT limits of $\CN=4$ SYM are dual to limits of type IIB strings on AdS$_5\times S^5$ with $E-J$ and $g_s$ going to zero with $(E-J)/ g_s$ and $N$ kept fixed. The effective string tension in AdS$_5\times S^5$ is 
\begin{equation}
\label{tension}
T = \frac{1}{2\pi} \sqrt{4\pi g_s N} \, , 
\end{equation}
which goes to zero in the SMT limits.

Four of the bounds do not involve all of the five angular momenta. Let $n$ denote the number of angular momenta not included in the bound. In the SMT limit the $2n$ directions - here called {\sl external directions} - that realize the rotation planes for these $n$ angular momenta have a confining potential with effective mass proportional to $1/g_s$ and hence these directions are forced to sit at the minimum of the potential. This gives an effective reduction of the number of spatial dimensions after the limit.

One can show that $\ads_5\times S^5$ admits a coordinate system $u$, $x^\mu$, $y^I$ where $\mu=0,1,2,...,d$, $d=8-2n$, and  $I=1,2,...,2n$, with the properties that i). $y^I$ are the $2n$ external directions that are confined to be at $y^I=0$ in the limit, ii). $\partial_u$ and $\partial_{x^0}$ are Killing vector fields with $i\partial_{x^0} = E-J$ and iii). the metric of AdS$_5\times S^5$ can be put in the form \eqref{eq:nullred} when setting $y^I=0$, with $\tau_\mu$, $m_\mu$ and $h_{\mu\nu}$ such that $\tau_0=1$ and $m_0=h_{00}=h_{0i}=0$ for $i=1,2,..,d$. 

The scaling limit introduced above corresponds to the SMT limit if one identifies $c^{-2} = 4\pi g_s N$. Following this, one rescales $x^0= c^2 \tilde{x}^0$ such that the rescaled energy $i \partial_{\tilde{x}^0} = (E-J)/ (4\pi g_s N)$ is kept fixed in the limit. The rescaled tension is $\tilde{T} = cT =  \frac{1}{2 \pi}$. After the scaling limit we get the action \eqref{eq:tildeS}. With the gauge choice $\sigma^0 = \frac{1}{P^2} X^0$ (with $X^0=\tilde{x}^0$ on the world-sheet) and $\tilde\eta=P \sigma^1$ this becomes \eqref{eq:Stildegauge0}. 

We conclude that the SMT limit applied on type IIB strings on AdS$_5\times S^5$ realizes the scaling limit $c\rightarrow \infty$ introduced above, and therefore corresponds to a non-relativistic limit both on the target space, as well as on the world-sheet. After the limit, the target space is a $d+1$ dimensional $U(1)$-Galilean geometry and the world-sheet theory is a non-relativistic two-dimensional theory. Note that the action \eqref{eq:Stildegauge0} is large if $P$ is large, and one can thus take a classical limit of the action, even if the SMT/scaling limit involves sending the effective tension $T$ to zero \cite{Harmark:2008gm}. See \cite{Harmark:2008gm} for a discussion of quantum effects in such limits.

\noindent\textbf{Examples} 

As the simplest example, consider the SMT/scaling limit towards the BPS bound $E\geq J=J_1+J_2$. Write the metric of AdS$_5\times S^5$ as 
\begin{eqnarray}
\label{eq:adsmet1}
&& g_{MN} dx^M dx^N =\cos^2 \psi[  2 \tau ( du - m ) + h_{\mu\nu} dx^\mu dx^\nu ]
 \nn \\ && - (\sinh^2 \rho + \sin^2 \psi) (dx^0-\frac{1}{2} du)^2 \nn \\ &&+ 
 d\rho^2+ \sinh^2\rho \, d\Omega_3^2 + d\psi^2 + \sin^2 \psi \, d\alpha^2 \,,
\end{eqnarray}
with $d=2$ since $n=3$, $\tau= dx^0 - 2m$ and
\begin{equation}
\label{eq:LLmh}
m = - \frac{\cos \theta}{2}  d\phi  \spa h_{\mu\nu}dx^\mu dx^\nu = \frac{1}{4} ( d\theta^2 + \sin^2 \theta d\phi^2 )  \,.
\end{equation}
Note that the radius is set to one and instead included in the tension \eqref{tension}. The six external directions have a potential proportional to $(\sinh^2\rho + \sin^2\psi)/g_s$ that confines them to the point $\rho=\psi=0$ \cite{Harmark:2008gm}. The SMT limit leads then to the 2+1 dimensional $U(1)$-Galilean geometry given by $\tilde{\tau} = d\tilde{x}^0$ and Eq.~\eqref{eq:LLmh}. The non-relativistic sigma-model \eqref{eq:Stildegauge0} is the Landau-Lifshitz model with $P=J$. Thus, we get a new interpretation of the Landau-Lifshitz model as a non-relativistic string theory of the form \eqref{eq:tildeS} with a $U(1)$-Galilean target space geometry.

SMT becomes a nearest-neighbor spin chain for $N=\infty$, which   is the ferromagnetic Heisenberg spin chain with $SU(2)$ symmetry
for $J=J_1+J_2$. In a long-wave length approximation with large $J$ this is described by the Landau-Lifshitz model hence matching the SMT/scaling limit on the string theory side. 

The connection between the emerging sigma-models from spin chains and limits of strings on AdS$_5 \times S^5$ was first pointed out in \cite{Kruczenski:2003gt} by Kruczenski and later studied for other sectors in \cite{Kruczenski:2004kw,Hernandez:2004uw,Stefanski:2004cw,Bellucci:2004qr}. These cases can all be interpreted in the framework of this paper as well. However, the Kruczenski limit does not correspond to our scaling limit since it does not take the tension \eqref{tension} to zero. Instead, it takes $J=J_1+J_2$ to infinity keeping $T^2/J$ fixed \cite{Kruczenski:2004kw}, hence it includes terms of higher orders in $T^2/J$ in contrast with the SMT limit. Moreover, one is in different regimes on the gauge theory and string theory sides. 

Another example is the limit towards the BPS bound $E\geq J=S_1+S_2+J_1+J_2+J_3$. Write the metric of 
AdS$_5\times S^5$ as 
\begin{eqnarray}
\label{eq:adsmet2}
& g_{MN} dx^M dx^N = - \cosh^2 \rho \, dt^2 +d\rho^2+ \sinh^2\rho\, d\Omega_3^2 + d\Omega_5^2\,, & \nn \\
& d\Omega_{2k+1}^2 = (d\Sigma_k)^2 + (d \chi_k + A_k)^2\,, &
\end{eqnarray}
where $E=i\partial_t$, $S_1+S_2= -i \partial_{\chi_1}$, $J_1+J_2+J_3=-i\partial_{\chi_2}$, $(d\Sigma_k)^2$ is the Fubini-Study metric on $\C P^k$ and $A_k$ is a one-form on $\C P^k$, $k=1,2$. Using $t = v - \frac{1}{2} u$, $\chi_1 = v - \frac{1}{2} u + w$ and $\chi_2 = v + \frac{1}{2} u$ the metric is of the form \eqref{eq:nullred} for $d=8$ with
\begin{equation}
\label{eq:su123_m}
\begin{array}{c} \ds
m = - \sinh^2 \rho ( dw + A_1 ) - A_2 \,, 
\\[2mm]
%\label{eq:su123_h}
\begin{array}{rcl} \ds
h_{\mu\nu}dx^\mu dx^\nu &=& d\rho^2+  \frac{1}{4} \sinh^2 ( 2\rho) (dw + A_1)^2 \\[2mm] && \ds + \sinh^2\rho \, d\Sigma_1^2 + d\Sigma_2^2 \,,
\end{array}
\end{array}
\end{equation}
and $\tau =  dx^0 + \frac{1}{2} m +A_2$. Taking the scaling limit gives now the $8+1$ dimensional $U(1)$-Galilean geometry defined by $\tilde{\tau}=d\tilde{x}^0$ and \eqref{eq:su123_m} with sigma-model given by \eqref{eq:tildeS} and  
 \eqref{eq:Stildegauge0}. This limit is of particular interest since it corresponds to the highest possible dimension of the target space, and the largest global symmetry group $SU(1,2|3)$ of the corresponding SMT and spin chain.

\noindent\textbf{Discussion} 

The results of this letter open up for a wide scope of directions. It would be worthwhile to understand better the nature of the $U(1)$-Galilean target space geometry.  Another important problem is to consider the quantum theory of the non-relativistic string actions \eqref{eq:NRstringTNC} and \eqref{eq:tildeS} that we have found, including beta-functions and the dynamical role played by the target space dimension
(for which we naturally get $d+1=3,5,9$ in the case of the limits on $AdS_5\times S^5$). 
In particular, since dynamical NC geometry is related to Ho\v rava--Lifshitz gravity \cite{Hartong:2015zia,Hartong:2016yrf}, 
it would be interesting to see if the couplings to the target space objects $\tau_\mu$, $m_\mu$, $h_{\mu\nu}$ in \eqref{eq:NRstringTNC} and $\tilde\tau_\mu$, $m_\mu$, $h_{\mu\nu}$ in \eqref{eq:tildeS} have to obey certain consistency conditions that can be interpreted as the equations of motion of a  non-relativistic gravity. For SMT, this could in turn be interesting since one should then be able to see the emergence of $U(1)$-Galilean geometry and its associated gravitational dynamics from a quantum theory.

Important generalizations and extensions of our results are: i). the effect of adding the NSNS $B$-field to the limits, which could be useful to understand if there is a notion of T-duality and if there is a relation with the Gomis--Ooguri formulation of non-relativistic closed strings \cite{Gomis:2000bd}, ii). the inclusion of fermions, and corresponding supersymmetric versions of the non-relativistic sigma-models and iii). a systematic study of higher derivative corrections to the sigma-models. Moreover, by  applying similar limits to the DBI D-brane action (see also \cite{Harmark:2016cjq})  it seems very likely that higher-dimensional non-relativistic world-volume theories should exist. 

\noindent\textbf{Acknowledgements.}
We thank Eric Bergshoeff, Jan de Boer, Diego Hofman and Gerben Oling  for useful discussions. The work of TH and NO is supported in part by FNU grant number DFF-6108-00340 and the Marie-Curie-CIG grant number 618284.

\addcontentsline{toc}{section}{References}
%\begin{thebibliography}{18}

%\bibliography{NC}

\end{document}